\documentclass [prl,twocolumn,superscriptaddress,showpacs] {revtex4}
\usepackage{graphicx,amsmath,amssymb,epsf,longtable}

\begin{document}

\title{
Collinear versus non-collinear magnetic order in Pd atomic clusters
}
\author{F. Aguilera-Granja}
\affiliation{Departamento de F\'{\i}sica Te\'orica, At\'omica
y \'Optica. Universidad de Valladolid, E-47011 Valladolid, Spain}
\author{J. Ferrer}
\affiliation{Departamento de F\'{\i}sica.  Universidad de Oviedo, Spain}
\author{A. Vega$^1$}

\date{\today}
\begin{abstract}
We present a thorough theoretical assessment of the stability of non-collinear 
spin arrangements in small palladium clusters. We generally find that 
ferromagnetic order is always preferred, but that antiferromagnetic and
non-collinear configurations of different sorts exist and compete for the 
first excited isomers. We also show that the ground state is insensitive to the 
choice of atomic configuration for the pseudopotential used and to the approximation 
taken for the exchange and correlation potential. Moreover, the existence and
relative stability of the different excited configurations also depends
weakly on the approximations employed. These results provide strong evidence on the 
transferability of pseudopotential and exchange and correlation functionals for
palladium clusters as opposed to the situation found for the bulk phases of palladium.
\end{abstract}


\maketitle

\section{Introduction}
The magnetic properties of free-standing atomic clusters of 3d TM 
elements have been intensively scrutinized during the last two decades. 
Two different but related phenomena have specifically been discussed 
and essentially unravelled. The first is the modification of local 
magnetic moments as compared with the values found in bulk materials. 
The second is the competition between the possible ferromagnetic, 
antiferromagnetic and non-collinear arrangements of the local spins, 
as well as its interplay with the geometry of the nanostructure. 
In the case of ferromagnetic elements like Fe,
Co and Ni, the increase of the average cluster magnetic moment can be
easily explained in terms of the reduced atomic coordination in the
low-dimensional regime, with oscillations associated to structural
(symmetry) changes. \cite{fagni}
The case of antiferromagnets like Cr and Mn is much more complex. Atoms 
of these elements may display large magnetic moments, since they have
a large number of d-holes susceptible to be polarized. On the other hand,
clusters of these atoms may display tiny average magnetizations due to the
tendency of their atomic moments to align in antiparallel directions.
The structure plays also a fundamental role in the magnetic behavior of these
clusters, since it may originate magnetic frustration. 
A conventional example of magnetic frustration in a classical spin system appears 
when atoms positions form triangular motifs. The studies of these classical
systems show that magnetic frustration frequently leads to non-collinear 
configurations of the local spin moments.
The latest theoretical studies reported in the literature show that non-collinear 
arrangements of quantum spins also appear as the ground or as some of the first 
isomers of clusters of $3d$ atoms, including not only Cr and Mn, but also Fe, 
Co and Ni.\cite{oda, kohl, hobbs, fujima, longo}

All materials made of $4d$ TM elements are paramagnets, in contrast to those
of the $3d$ row. A natural question thus arises of whether small clusters of 
$4d$ elements may show low-lying magnetic states of collinear or even 
non-colinear nature. Bulk palladium, being a paramagnet in the brink of becoming
a ferromagnet, presents one of the most intriguing and controversial 
magnetic behaviors in nature.\cite{Chen} It is therefore not surprising that 
the very few experimental and theoretical studies published so far try to 
clarify whether Pd clusters of given sizes are magnetic or not, and what is the 
order of magnitude of their average magnetic moment. From the experimental side, 
most of the reports agree that only very small clusters have a net magnetic 
moment \cite{Douglass, Cox, Taniyama, Sampedro}, with the exception of 
Shinohara and coworkers, \cite{Shinohara} who found noticeable magnetic moments 
at the surface of Pd particles as big as 79 \AA. From the theoretical side, there 
is also consensus that very small Pd clusters are indeed 
magnetic.\cite{Lee,Vitos,Reddy,Moseler,Kumar,FAG_Pd}
Futschek et al.\cite{Hafner} have studied recently small Pd clusters 
using Density Functional Theory (DFT) in the collinear framework, 
within a fixed-moment mode.  They have found that multiple spin isomers 
exist for each cluster size with very small energy differences.  Interestingly,
some of these competing isomers present ferromagnetic order, while others display 
antiferromagnetic alignments, with possible frustration. Although Pd has tendency 
to ferromagentic order, this fact strongly points out to
the possible existence of non-collinear magnetic structures, as a mechanism
to release the frustration and competition between the different magnetic solutions.

We report in this article a thorough {\it Ab initio} study of the
magnetic behavior of small palladium clusters Pd$_N$, with $N$ ranging
from 3 to 7.
We have performed a simultaneous optimization of the
geometric and magnetic degrees of freedom fully allowing for
non-collinear spin arrangements.  A debate exists currently on the accuracy of the
Local Density Approximation (LDA) \cite{CA} versus the Generalized
Gradient Approximation (GGA) \cite{PBE} for the determination of the
magnetic behavior of low-dimensional Pd systems
\cite{Reddy,Moseler,Kumar,Delin,Alexandre}. The present article shows
that both approximations provide the same results for the ground state
of free-standing Pd atomic clusters. Moreover, even the existence and
energy ordering of the first excited states depend much more weakly on
the choice of correlation functional and pseudopotential in than the bulk
fcc material.

\begin{figure} \centerline {\includegraphics[angle=-00.0,
width=0.95\linewidth] {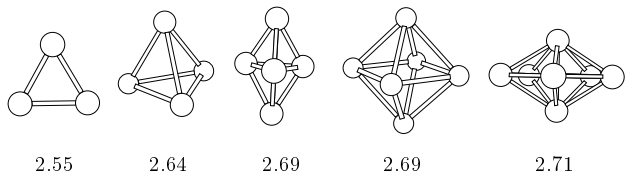}} \vspace{-1pt}
\caption{Illustration of the ground state structures of the different 
clusters here studied and average interatomic distances (in \AA) 
within GGA1.}  \label{Fig1} \end{figure}

\section{Theoretical details}

We have performed our calculations using the code SIESTA.\cite{SIESTA}
SIESTA is a DFT method that employs linear combination of pseudoatomic
orbitals as basis set. The electronic core is replaced by a nonlocal
norm-conserving Troullier-Martins\cite{TM_1991} pseudopotential that may
include nonlinear core correction terms.  The code allows to
perform, together with the electronic calculation, structural
optimization using a variety of algorithms. It also allows to 
simulate non-collinear spin arrangements both in the LDA and in the 
GGA approximations. \cite{Suarez} 

In the present calculation, we have also used a variety of pseudopotentials
to test their effect on free-standing clusters and their corresponding 
transferability. We have generated three different pseudopotentials 
using LDA. The first (LDA1) was built with the electronic configurations 
$5s^1$, $5p^0$ and $4d^9$, and core-corrections matching radius $r_c=2.00$ a.u.; the
second (LDA2) was identical to LDA1, but with $r_c=1.2$ a.u.; the third
had a closed-shell atomic configuration ($5s^0$, $5p^0$ and $4d^{10}$) and 
$r_c=1.2$ a.u. We have also generated two GGA pseudopotentials with electronic
configuration $5s^1$, $5p^0$ and $4d^9$, and $r_c=2.0$ or 1.2 a.u.
(GGA1 and GGA2, respectively).  In all five cases, the cutoff radii of the s, p 
and d orbitals were taken at 2.30, 2.46 and 1.67 a.u., respectively.
We have described valence states by a double-$\zeta$ polarized basis set (e.g.:
two different radial functions for s and d orbitals and a single one for p orbitals). 
We have taken an energy cutoff of 150 Ry to define the real space grid for numerical 
integrations, but we checked that higher cutoffs did not alter the results.
We have carried out the structural optimization using a conjugate gradient algorithm, 
where we have set the tolerance for the forces at 0.003 eV/\AA, with eventual 
double-checks using 0.001 eV/\AA.

\section{Results}

We have found that the five pseudopotentials provide similar results
when applied to an isolated palladium atom, being the eigenvalues of
the ground state and different excited states slightly better
reproduced with LDA1 and GGA1 (both had $r_c=2.00$ a.u.). However, we
have observed that they give rise to different magnetic behaviors when
applied to the bulk fcc material. We first remind that palladium has a
paramagnetic ground state with a lattice constant of 3.89 \AA.
Moreover, the latest simulations show that LDA predicts a paramagnetic
ground state with a lattice constant of about 3.85-3.91 in agreement
with experiments, while GGA predicts the ground state to be
ferromagnetic, with a much larger lattice constant of about 4.0 \AA \,
and a magnetic moment of about 0.4 $\mu_B$. The fact that LDA provides
a slightly better description of the ground state of bulk hcp has been
stressed recently by Alexandre and co-workers \cite{Alexandre}. Our
LDA1 pseudopotential gives a ferromagnetic ground state with $M
\approx 0.54 \mu_B$, while LDA2 and LDA3 predict the ground state to
be paramagnetic, as it should. Finally, both GGA pseudopotentials lead
to a ferromagnetic ground state with $M \approx 0.48 \mu_B$.  On the
other hand, all LDA approximations give a lattice constant equal to
3.90 \AA.  In contrast, both GGA pseudopotentials predict it to be
equal to 4.01 \AA.  We note here that the fact that all LDA
pseudopotentials predict the same lattice constant for the ground
state regardless of its magnetic nature also happens with fcc Iron,
where LDA predicts the same ground state energy and lattice constant
for the paramagnetic, antiferromagnetic and ferromagnetic Low Spin
states \cite{Moruzzi,Suarez}.  These results highlight the lack of
transferability of the different pseudopotential and the sensitivity
with the exchange and correlation functional used to describe the ground
state of bulk palladium.

It is therefore of great importance to assess whether the same
difficulties apply to the case of the atomic clusters considered here.
As we shall discuss below, atomic palladium clusters are much more
insensitive to the choice of pseudopotential and exchange and
correlation functional. We shall show that the ground state is the
same for the different approximations used here, in stark contrast to
the case of the bulk material. Further, a large portion of the
low-lying excited states are reproduced by all approximations.

\begin{table} 
\caption{Bindig energy of the ferromagnetic clusters in meV/atom.}  
 \vspace{-7pt}
\begin{center}
\small{\renewcommand{\arraystretch}{0.10}\renewcommand{\tabcolsep}{0.35pc}
\begin{tabular}{lcccccc}
\hline
 N   & \, LDA1 \,  & \, LDA3 \, & \, GGA1  \, &  \,Ref.[\onlinecite{Kumar}] \, & \,
Ref.[\onlinecite{Hafner}] \,  \\
\hline
3  &   1.755  & 1.326  &  1.289 &  1.203  & 1.250  \\
4  &   2.293  & 1.942  & 1.769  &  1.628  & 1.675  \\
5  &   2.502  &  2.168 & 1.933  &  1.766  & 1.805  \\
6  &   2.721  &  2.401   &  2.110&  1.919  & 1.949  \\
7  &   2.791  &  2.452   &  2.155 &  1.953  & 1.985  \\
\hline
\end{tabular}}
\end{center}
\end{table}

Notice that we have not kept fixed the magnetic moment in our simulations
of the Pd$_N$ clusters, but rather have allowed it to vary freely during 
the non-collinear iterative selfconsistency process, in contrast to previous
authors. Moreover, while we can not rule out that we may have missed
low lying solutions, we have endeavored to minimize this risk
by feeding a large variety of non-collinear seeds for each cluster.
This effort has allowed us to find a rich and complex family of
metastable solutions, that was absent in previous works. We finally
note that we have repeated all calculations with the pseudopotentials
LDA1, LDA3 and GGA1.

We have found that all clusters, except Pd$_6$, share the same
collinear magnetic ground state, with a total spin of 2 $\mu_B$, in
agreement with previous authors \cite{Kumar,Hafner}.  The geometry of
the ground state and the average interatomic distance of the Pd$_N$
clusters is displayed in Fig. 1, where we show that these range from
2.55 \AA \, in ${\rm Pd}_3$ to 2.71 \AA \, for ${\rm Pd}_7$. We have
written the binding energies of the different clusters in Table I. The
table shows that GGA1 gives slightly smaller values than LDA1 and
LDA3, as otherwise expected.  Moreover, the binding energies predicted
by GGA1 are very similar to those obtained by Kumar, who also used the
GGA (within an ultrasoft pseudopotentials, plane waves code) and by
Futschek et al., who used the all-electron VASP code, but did not
state the approximation employed.

We should stress that all the tested pseudopotentials provide the same
ground state, in stark contrast to the situation that arose for the bulk
material. Moreover, we have found very similar inter-atomic distances for
all Pd$_N$ clusters, as shown in Table II.  The table shows that GGA1 predicts 
slightly longer distances than LDA1 or LDA3, as otherwise. These distances also
agree with those obtained by Kumar and Futschek within a range of 1 per cent.

\begin{table} 
\caption{Inter-atomic distances of the ferromagnetic clusters in Angstrom.}  
 \vspace{-7pt}
\begin{center}
\small{\renewcommand{\arraystretch}{0.10}\renewcommand{\tabcolsep}{0.35pc}
\begin{tabular}{lcccccc}
\hline
 N   & \, LDA1 \,  & \, LDA3 \, & \, GGA1  \, &  \,Ref.[\onlinecite{Kumar}] \, & \,
Ref.[\onlinecite{Hafner}] \,  \\
\hline
3  &   2.49  &  2.49  &  2.56  &  2.52  & 2.52  \\
4  &   2.57  &  2.59  &  2.64  &  2.61  & 2.61  \\
5  &   2.61  &  2.63  &  2.70  &  2.65  & 2.64  \\
6  &   2.62  &  2.64  &  2.70  &  2.66  & 2.66  \\
7  &   2.64  &  2.66  &  2.72  &  2.68  & 2.70  \\
\hline
\end{tabular}}
\end{center}
\end{table}

\begin{figure}[b] \centerline {\includegraphics[angle=-90.0,
width=0.99\linewidth] {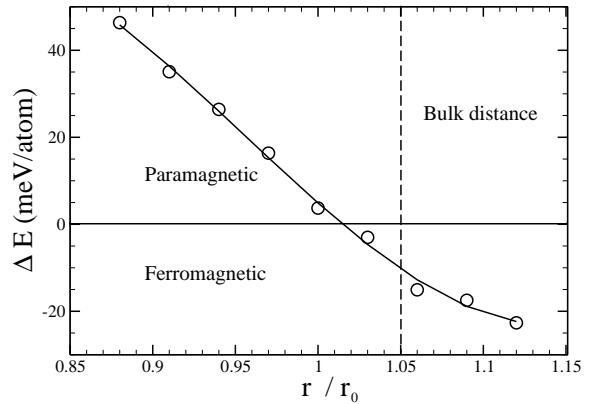}} 
\caption{Energy difference of the Ferromagnetic state and the
Paramagnetic one for the Pd$_6$ clusters as a function of 
the average inter-atomic distance. We have taken the energy of the paramagnetic ground
state as the zero of energy. Distances are measured in units of the equilibrium distance
of the cluster. The vertical dashed  line indicates the theoretical interatomic distance
of bulk palladium in the GGA1 approximation.}  \label{Fig1b} 
\end{figure}

\begingroup
\squeezetable
\begin{table*}
\caption{Different solutions obtained for the Pd$_N$ clusters. We provide the absolute
values of the atomic magnetic moments, the total magnetic moment in the cluster $\bar\mu$
(both in units of $\mu_B$) and the excitation energy per atom (in meV). For $N$=5 and 7, the
first two values of the atomic moments correspond to the axial sites, whereas the last ones 
correspond to the planar sites.}
\medskip
\begin{center}
\small{\renewcommand{\arraystretch}{0.75}\renewcommand{\tabcolsep}{0.001pc}
\begin{tabular}{lcccccccccc}
\hline
\hline
    &LDA1 &  &  & LDA3 &  & & GGA1  & &   \\
\hline
    & Local moments &  $ \bar \mu$ \,\, & $ \Delta E$  & Local moments &  \, $ \bar \mu$ \, & \, $ \Delta E$ \, &
Local moments & \, $ \bar \mu$ \, \, & $ \Delta E$  & \\
\hline
N=3 & & & & & & & & & \\ 
\hline
Ferro. & (0.67$\times$3)& {\bf 2}   & 0 
       & (0.67$\times$3)  & {\bf 2} & 0 
       & (0.67$\times$3)  & {\bf 2} & 0 \\ 
AF     &  &  &   
       &  &  & 
       &  (0, 0.30, -0.30) & {\bf 0} & 28  \\
Radial &  &   &  
       & &  & 
       & (0.18$\times$3) &  {\bf 0} & 28   \\
Para.  &  (0$\times$3) &  {\bf 0}  & 68  
       &  (0$\times$3) & {\bf 0} &  34  
       &  (0$\times$3) & {\bf 0} &  75  \\
\hline
N=4 & & & & & & & & & \\ 
\hline
Ferro. & (0.50$\times$4)& {\bf 2} & 0 &(0.50$\times$4)  & {\bf 2} &0 & 
(0.50$\times$4)& 
{\bf 2} & 0 \\ 
NC1  &  & &  & & &  & (0.29,0.29,0.29,0.29) &{\bf 0} & 12 \\
NC2  & (0.35,0.24,0.24,0.35) & {\bf 0.25} & 9 &(0.25,0.28,0.28,0.25) & {\bf 0.03}  & 10 &&&   \\
AF1  & (0.32, 0.32, -0.32, -0.32) & {\bf 0} & 26 & 
(0,23, 0.23, -0.23, -0.23) & {\bf 0}  & 30 
& (0.29, 0.29, -0.29, -0.29) & {\bf 0} & 25 \\    
AF2 & (0.41, 0, -0.41, 0) &  {\bf 0} & 40 
&  (0,31, 0, -0.31, 0) &  {\bf 0} & 31 
&  (0.38, 0, -0.38, 0) &{\bf 0}  & 36 \\  
Para.    &  (0$\times$4) & {\bf 0} &  86  &  (0$\times$4) & {\bf 0} & 59 &  (0$\times$4) & {\bf 0} & 78 \\
\hline
N=5 & & & & & & & & & \\ 
\hline
Ferro. & (0.43,0.43,0.38$\times$3)& {\bf 2} & 0 
       & (0.40$\times$5)& {\bf 2} & 0 
       &(0.42,0.42,0.39$\times$3)  & {\bf 2} &0 \\ 
AF1 & (0, 0, 0.43, - 0.43, 0)& {\bf 0} & 22 
    &  (0, 0, 0.33, - 0.33, 0)&  {\bf 0}  & 19 
    &  (0, 0, 0.39, - 0.39, 0)&  {\bf 0}  & 18 \\
AF2  &  (0, 0, 0.48, -0.24, -0.24)& {\bf 0} & 27
     &  & &  
     & (0, 0, 0.44, -0.22, -0.22)& {\bf 0} & 19  \\ 
Radial & (0, 0, 0.29$\times$3)& {\bf 0} & 35  
       & & &
       &  (0, 0, 0.27$\times$3)& {\bf 0} & 28 \\ 
Para.      &  (0 $\times$5) & {\bf 0}  & 63  
           &  (0  $\times$5) & {\bf 0}  & 41 
           &  (0 $\times$5) & {\bf 0}  & 55\\ 
\hline
N=6 & & & & & & & & & \\ 
\hline
Ferro. & (0.33$\times$6)& {\bf 2} & 0 
       & (0.33$\times$6)& {\bf 2} & 0 
       & (0.33$\times$6) & {\bf 2} &0 \\  
Para.  &  (0$\times$6) &{\bf 0} & -\,13  
       & (0$\times$6) & {\bf 0} & -\,12   
       &  (0$\times$6) & {\bf 0} & -\,4 \\ 
\hline
N=7 & & & & & & & & & \\ 
\hline
Ferro. & (0.19,0.19,0.32$\times$5)& {\bf 2} & 0 
       & (0.21,0.21,0.31$\times$5)& {\bf 2} & 0
       & (0.20,0.20,0.32$\times$5)& {\bf 2} & 0 \\ 
AF1 & (-0.36,0.36,-0.33,-0.22,0.22,0.32,0)&  {\bf 0} & 9
    &                           &                &     
    & (-0.32,0.32,-0.30,-0.20,0.20,0.30,0)&  {\bf 0} & 8   \\ 
AF2 &  (0,0,-0.36,-0.23,0.23,0.36,0)& {\bf 0} & 14
    & \, (0,0,-0.29,-0.20,0.20,0.29,0)& {\bf 0} & 8  
    &  (0,0,-0.32,-0.21,0.21,0.32,0)& {\bf 0} & 12   \\
Radial  & (0.27,0.27,0.18$\times$5) & {\bf 0} & 22   
        & (0.24,0.24,0.12$\times$5) & {\bf 0} & 14 
        & (0.24,0.24,0.17$\times$5) & {\bf 0} & 20 \\
Para.   &  (0$\times$7) & {\bf 0} & 37  
        & (0$\times$7) & {\bf 0} & 24  
        & (0$\times$7) & {\bf 0} & 33 \\ 
\hline
\hline
\end{tabular}}
\end{center}
\end{table*}
\endgroup

The Pd$_6$ cluster displays a behavior different from the rest, and
therefore we discuss it separately. Futschek and coworkers
\cite{Hafner} found that Pd$_6$ was also ferromagnetic in contrast to
Kumar et al.\cite{Kumar}, who predicted it to be paramagnetic. We have
found that both states are nearly degenerate, with the paramagnetic
solution being slightly more stable. Further, we have found that the
relative energy of the two solutions show a strong dependence with
inter-atomic distance.  Fig. (2) shows that there is a level crossing
when the bonds are elongated by just 1 per cent from the equilibrium
distance. The close proximity between the equilibrium distance and the
level crossing distance explains the sensitivity of these results with
respect to slight differences in the simulations and the discrepancies
between Kumar and Futscheck \cite{Kumar,Hafner}.  Aditionally, we have
been unable to find non-collinear or antiferromagnetic solutions for
this cluster.

\begin{figure}[b] \centerline {\includegraphics[width=0.9\linewidth,angle=0] {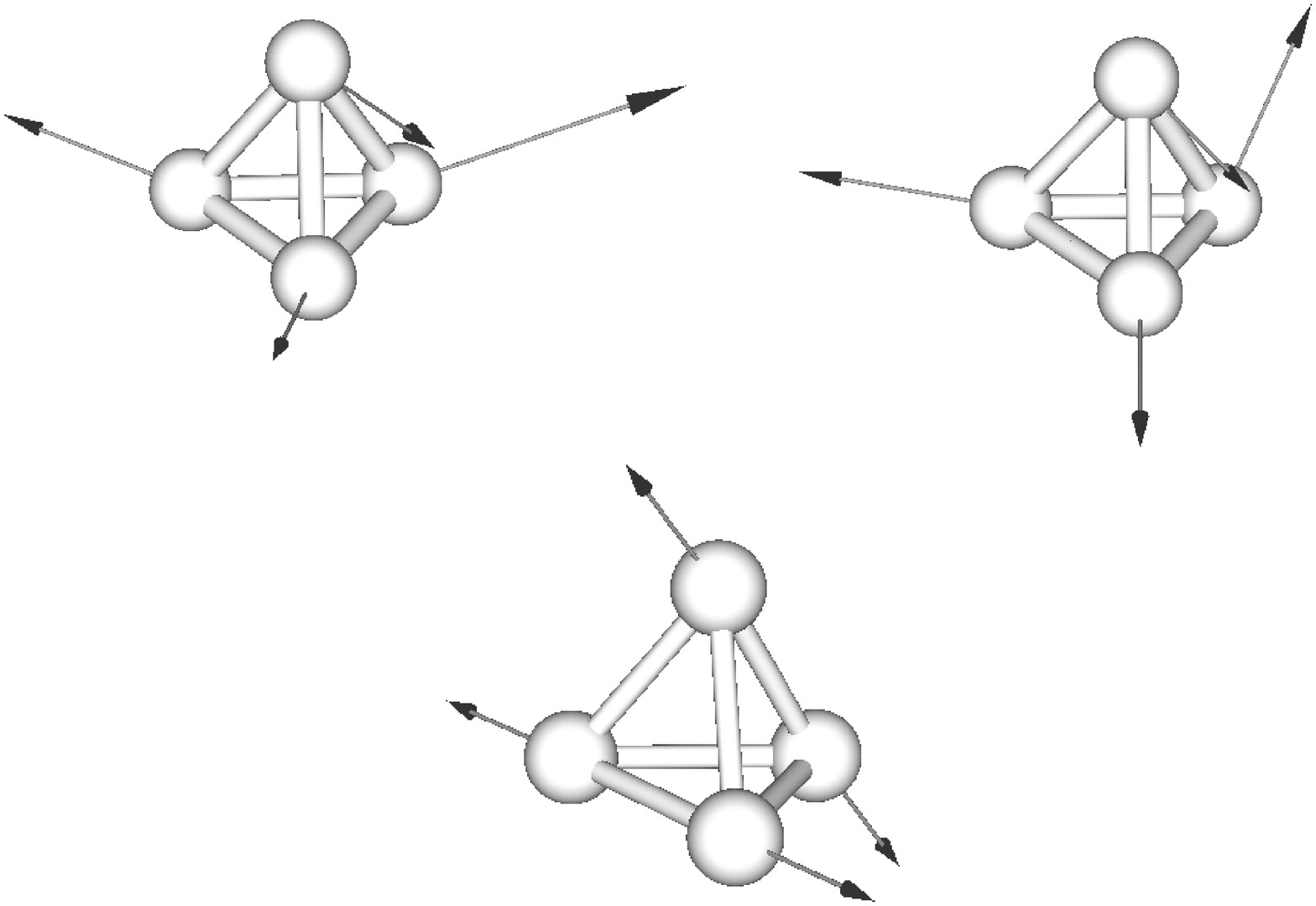}} 
\vspace{20Pt}
\caption{Illustration of the non-colinear magnetic solutions 
for ${\rm Pd}_4$ NC2 (LDA1), NC2 (LDA3) and NC1 (GGA1). The arrows are proportional 
to the size of the atomic moments.}  \label{Fig2} \end{figure}

In contrast, and independently of the pseudopotential or approximation used, 
the rest of the clusters show a rich variety of antiferromagnetic and non-collinear 
solutions. Most of these solutions, though not all, exist for all LDA1, LDA3 and GGA1.
We have also found that, whenever they exist, the relative order of the different
solutions is maintained, and the size of the atomic moments is very similar. These facts 
strengthen our belief that Pd atomic clusters are much more insensitive to the 
pseudopotential and approximation employed than bulk Pd. It is also reassuring
that most of the collinear solutions have been identified in previous 
calculations\cite{Hafner} (e.g.: AF1 for Pd$_4$ and Pd$_5$ and AF2 for Pd$_7$).

\begin{figure} \centerline{\includegraphics[height=9cm,width=7cm,angle=-90]{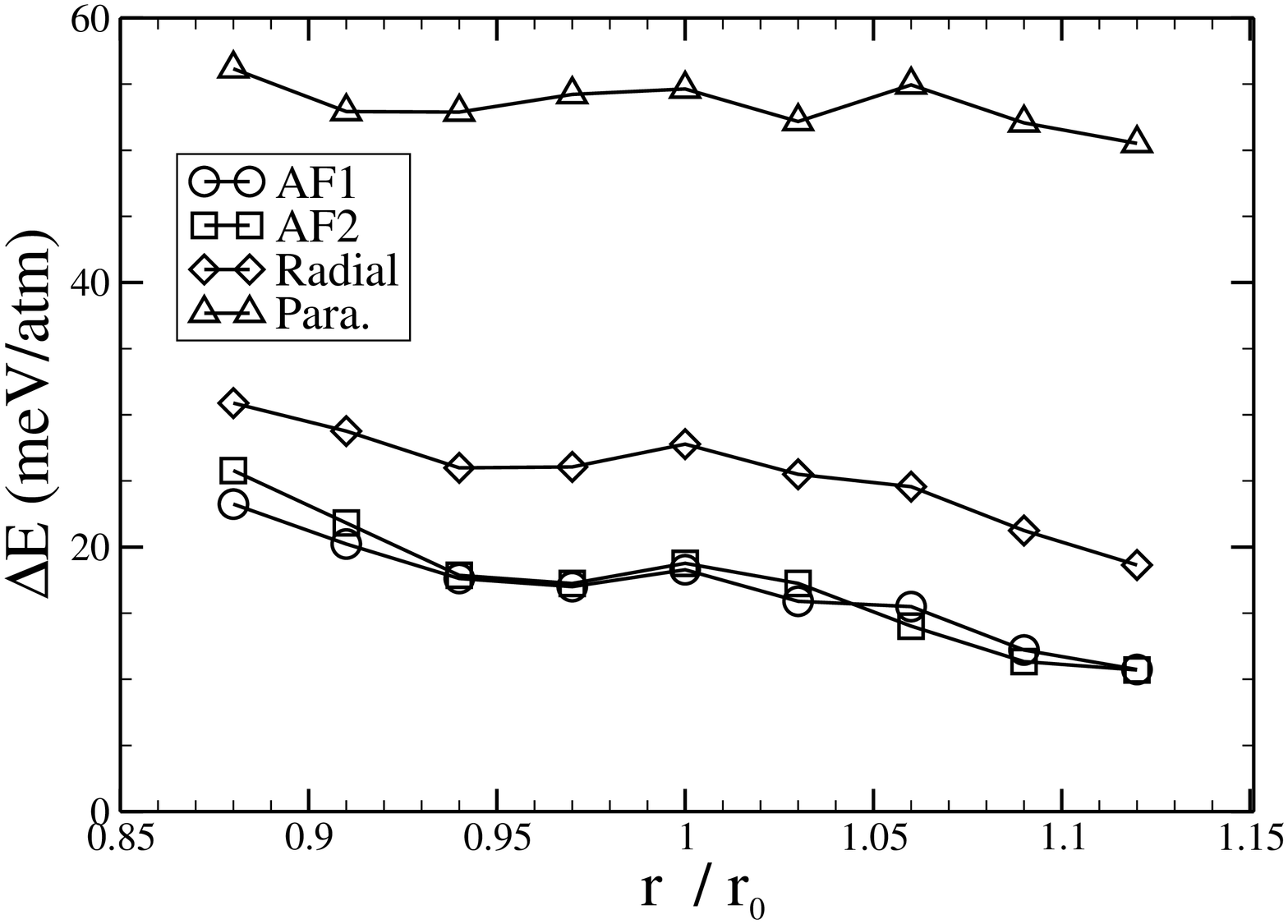}}
\caption{Excitation energy per atom of the magnetic solutions of the Pd$_5$ cluster as a 
function of the average interatomic distance, using GGA1.} \label{Fig3} \end{figure}

The non-collinear solutions found can be classified into those that
release antiferromagnetic frustration and therefore have lower
excitation energy than the AF solution (NC1 and NC2 in Pd$_4$, shown
in Table III and Fig. 3), and radial or quasi-radial solutions, that
resemble the hedgehogs found in low dimensional theories of classical
or quantum antiferromagnets\cite{Haldane}. Hedgehogs in these theories
do not release frustration but rather are excitations over the
antiferromagnetic ground state. We also find that these radial states
have a higher energy that the antiferromagnetic solution, and
therefore do not release frustration.

Notice that the antiferromagnetic and non-collinear solutions can be
reached at temperatures of the order of room temperatures (25 meV).
Therefore, any measurement of the magnetization performed at room 
temperature should find a thermal average of all those states, many 
of which have a tiny magnetic moment. It should not be surprising that
such a measurement give a small net moment. 

We finally discuss the relationship between magnetism and equilibrium
interatomic distances. We have found that these are essentially the
same regardless of the magnetic state for the largest clusters ($N$= 5
- 7), the smallest ones showing slight variations of less than 0.04
\AA, but only within the LDA solutions. We have additionally analyzed
the relative stability of the different solutions as a function of the
interatomic distance. To this aim, we plot in Fig. 4 the energy per
atom of the low-lying excited states of the Pd$_5$ cluster, relative
to the ground state energy, as a function of an uniform volume
expansion, obtained using GGA1. The figure shows that no crossover
takes place, apart from the nearly-degenerate AF1 and AF2 solutions,
that cross at an expansion of about 4\%. Moreover, the relative energy
differences are essentially preserved and the local magnetic moments
kept constant, except for the AF2 and radial solutions, where they
slightly change (by about 10\%).

\section{Conclusions}

To summarize, we have studied the geometry and magnetic properties of the ground 
state and lowest lying isomers of small palladium clusters Pd$_N$, with $N$ ranging 
from 3 to 7. Our results confirm that the ground state is indeed collinear or 
paramagnetic. We have found a rich variety of non-collinear low-lying isomers, some 
of which efficiently release frustration, while other (hedgehog-like solutions) do not.
All these solutions should contribute to the room temperature magnetic behavior of the 
clusters, probably rendering small measured magnetic moments.
We have finally found that simulations of atomic clusters are rather insensitive to the
choice of the pseudopotential and to the approximation
used for the exchange and correlation potential.

To provide a rough explanation of this contrasting behavior, we first note that
GGA and LDA provide results that are quantitatively very similar in
most cases, including bulk palladium. GGA provides results that are
qualitatively different from LDA for bulk palladium because this
material sits right at the edge of the paramagnetic-ferromagnetic
transition, actually tilted slightly towards the paramagnetic side.
Therefore, bulk palladium falls slightly short of fulfilling Stoner's
criterion. GGA and LDA are mean field approximations. Hence, they tend to
overstimate magnetic properties. This error is partly compensated by
LDA, that predicts usually lower magnetic moments than GGA.
Consequently, LDA predicts the correct ground state for bulk palladium,
while GGA provides a ground state that is slightly tilted
towards the magnetic side of the phase diagram.
Hence, LDA and GGA predict quantitatively similar, but qualitatively
different results for the magnetic properties of this material.
Atomic Pd clusters are strongly magnetic, with the exception of Pd$_6$.
Therefore, all these approximations predict quantitative and qualitative
similar results for those clusters.

\section{Acknowledgments}

This work was supported by the Spanish Ministerio de Educacion y Ciencia 
(grants MAT2005-03415, BFM2003-03156 and SAB2004-0129)
in conjuction with the European Regional Development Found, INTAS (Project 03-51-4778)
and the Mexican (PROMEC-SEP-CA-230 and CONACyT-SNI). We wish to acknowledge useful 
conversations with J. M. Montejano-Carrizales, L. Fern\'andez-Seivane, 
V. M. Garc\'{\i}a-Su\'arez and F. Yndurain. 

{}

\end {document}